\newcommand{\extended}[1]{}    
\newcommand{\short}[1]{#1}     

\documentclass[runningheads]{llncs}

\usepackage{amsmath}
\usepackage{soul,color}
\usepackage{listings}
\usepackage{tikz}
\usepackage{graphicx}
\usepackage{epstopdf}
\usepackage{wrapfig}
\usepackage[caption=false,font=footnotesize]{subfig}
\usepackage{rotating}
\usepackage[normalem]{ulem}

\usepackage{wojtek11}
\usepackage{macros}

\usetikzlibrary{arrows,shapes,patterns,calc}
\tikzstyle{every node} = [draw = none, fill = white, thin]
\tikzstyle{every edge} += [black, thick]
\tikzstyle{noall} = [draw = none, fill = none]
\tikzstyle{nodraw} = [draw = none, fill = white]
\tikzstyle{nofill} = [draw = black, fill = none]
\tikzstyle{cnode} = [circle, draw = black]
\tikzstyle{snode} = [regular polygon, regular polygon sides = 4, draw = black]
\tikzstyle{lnode} = [diamond, draw = black]

\newcommand{\cmplx}[2][]{\mathit{compl}_{#1}(#2)}

\newcommand{\natatl}{\lan{NatATL}}
\newcommand{\ATL}{\lan{ATL}}

\newcommand{\thus}{\leadsto}
\newcommand{\gcmd}[2]{{#1}\thus{#2}}   
\newcommand{\size}{\mathit{length}}
\newcommand{\natstratsnorecl}[1][]{\Sigma_{#1}}
\newcommand{\Bool}{\mathcal{B}}

\newcommand{\Coercer}{Coercer\xspace}
\newtheorem{NatStr}{Natural Strategy}
\newcommand{\Selene}{\textbf{\sc Selene}\xspace}
\newcommand{\Uppaal}{\textsc{Uppaal}\xspace}
\newcommand{\Pret}{Pr\^et \`a Voter\xspace}

\newcommand{\vVote}{\text{vVote}\xspace}

\begin{document}

	\title{Natural Strategic Abilities in Voting Protocols}
	\author{Wojciech Jamroga\inst{1,2} \and
	Damian Kurpiewski\inst{2} \and
	Vadim Malvone\inst{3}}
	\institute{SnT, University of Luxembourg \and
Institute of Computer Science, Polish Academy of Sciences, Warsaw, Poland \and
Universit\'{e} d'\'{E}vry, France}

	\maketitle

\begin{abstract}
Security properties are often focused on the technological side of the system. One implicitly assumes that the users will behave in the right way to preserve the property at hand. In real life, this cannot be taken for granted. In particular, security mechanisms that are difficult and costly to use are often ignored by the users, and do not really defend the system against possible attacks.

Here, we propose a graded notion of security based on the complexity of the user's strategic behavior. More precisely, we suggest that the level to which a security property $\varphi$ is satisfied can be defined in terms of (a) the complexity of the strategy that the voter needs to execute to make $\varphi$ true, and (b) the resources that the user must employ on the way. The simpler and cheaper to obtain $\varphi$, the higher the degree of security.

We demonstrate how the idea works in a case study based on an electronic voting scenario. To this end, we model the vVote implementation of the \Pret voting protocol for coercion-resistant and voter-verifiable elections. Then, we identify ``natural'' strategies for the voter to obtain receipt-freeness, and measure the voter's effort that they require. We also look at how hard it is for the coercer to compromise the election through a randomization attack.


	\end{abstract}

\section{Introduction}\label{sec:intro}
Security analysis often focuses on the technological side of the system. It implicitly assumes that the users will duly follow the sequence of steps that the designer of the protocol prescribed for them. However, such behavior of human participants seldom happens in real life. In particular, mechanisms that are difficult and costly to use are often ignored by the users, even if they are there to defend those very users from possible attacks.

For example, protocols for electronic voting are usually expected to satisfy \emph{receipt-freeness} (the voter should be given no certificate that can be used to break the anonymity of her vote) and the related property of \emph{coercion-resistance} (the voter should be able to deceive the potential coercer and cast her vote in accordance with her preferences)~\cite{Benaloh94receipt\extended{,Okamoto98receipt},Juels05coercion,Delaune06coercion,Kusters10game\extended{,Dreier12formal}}.
More recently, significant progress has been made in the development of voting systems that would be coercion-resistant and at the same time \emph{voter-verifiable}, i.e., would allow the voter to verify her part of the election outcome~\cite{Ryan15verifiability,Cortier16sok-verifiability}.
The idea is to partly ``crowdsource'' an audit of the election to the voters, and see if they detect any irregularities.
Examples include the \Pret protocol~\cite{Ryan10atemyvote} and its implementation vVote~\cite{Culnane15vvote} that was used in the 2014 election in the Australian state of Victoria.

However, the fact that a voting system includes a mechanism for voter-verifiability does not immediately imply that it is more secure and trustworthy. This crucially depends on how many voters will actually verify their ballots~\cite{Verifiedvoting19policy}, which in turn depends on how understandable and easy to use the mechanism is.
The same applies to mechanisms for coercion-resistance and receipt-freeness, and in fact to any optional security mechanism.
If the users find the mechanism complicated and tiresome, and they can avoid it, they will avoid it.

Thus, the right question is often not \emph{if} but \emph{how much} security is obtained by the given mechanism.
In this paper, we propose a graded notion of \emph{practical security} based on the complexity of the strategic behavior, expected from the user if a given security property is to be achieved. More precisely, we suggest that the level to which property $\varphi$ is ``practically'' satisfied can be defined in terms of (a) the complexity of the strategy that the user needs to execute to make $\varphi$ true, and (b) the resources that the user must employ on the way. The simpler and cheaper to obtain $\varphi$, the higher the degree of security.

Obviously, the devil is in the detail. It often works best when a general idea is developed with concrete examples in mind. Here, we do the first step, and look how the level of coercion-resistance and voter-verifiability can be assessed in vVote and \Pret.
To this end, we come up with a multi-agent model of vVote, inspired by interpreted systems~\cite{Fagin95knowledge}.
We consider three main types of agents participating in the voting process: the election system, a voter, and a potential coercer. Then, we identify strategies for the voter to use the voter-verifiability mechanism, and estimate the voter's effort that they require. We also look at how difficult it is for the coercer to compromise the election through a randomization attack~\cite{Juels05coercion}.
The strategic reasoning and its complexity is formalized by means of so called \emph{natural strategies}, proposed in~\cite{Jamroga19natstrat-aij,Jamroga19natstratii} and consistent with psychological evidence on how humans use symbolic concepts~\cite{Bourne70concepts,Feldman00conceptlearning}.

To create the models, we use the \Uppaal model checker for distributed and multi-agent systems~\cite{Behrmann04uppaal-tutorial}, with its flexible modeling language and intuitive GUI. This additionally allows to use the \Uppaal verification functionality and check that our natural strategies indeed obtain the goals for which they are proposed.

\para{Related work.}
Formal analysis of security that takes a more human-centered approach has been done in a number of papers, for example with respect to insider threats~\cite{Hunker11insider-threats}. 
A more systematic approach, based on the idea of \emph{security ceremonies}, was proposed and used in~\cite{Carlos12ceremonies,Bella14concertina,Bella15servicesecurity,Martimiano16threatmodelling}, and applied to formal analysis of voting protocols~\cite{Martimiano15ceremony}. 
Here, we build on different modeling tradition, namely on the framework of \emph{multi-agent systems}. This modeling approach was only used in~\cite{Jamroga18Selene} where a preliminary verification of the \Selene voting protocol was conducted.
Moreover, to our best knowledge, the idea of measuring the security level by the complexity of strategies needed to preserve a given security requirement is entirely new.

Other (somewhat) related works include social-technical modeling of attacks with timed automata~\cite{David15sociotechnical-attacks} and especially game-theoretic analysis of voting procedures~\cite{Buldas07evoting,Culnane16benalohGT,Basin17eve,Jamroga17preventing}.
Also, strategies for human users to obtain simple security requirements were investigated in~\cite{Basin16humanerrors}.
Finally, specification of coercion-resistance and receipt-freeness in logics of strategic ability was attempted in~\cite{Tabatabaei16expressing}.


\section{Methodology}\label{sec:methodology}
The main goal of this paper is to propose a framework for analyzing security and usability of voting protocols, based on how easy it is for the participants to use the functionality of the protocol and avoid a breach of security.
Dually, we can also look at how difficult it is for the attacker to compromise the system.
In this section we explain the methodology\extended{, and describe the steps that we will take in the subsequent sections, from an informal analysis of the voting protocol, through creating a formal model of the voting system, to identifying the relevant strategies of the participants and measuring their complexity}.

\extended{
  \subsection{Preliminary Analysis of the Protocol}

  The first step is to choose an e-voting protocol. To make the analysis more grounded, we will analyze an actual implementation rather than the abstract description. Most scientific proposals of e-voting protocols omit many important steps and details, that only come out when the protocol is implemented and used in practice. In order to measure the complexity of strategies of the participants, we need to have a detailed, step by step instruction for the voter. Steps like logging in to the system, scanning one's ID card, checking the input etc. are very important when it comes to an assessing the complexity of a behavior.

  The next step is to choose the abstraction level for modeling. Usually a model of the whole procedure, including all the details, would be too complex for validation and analysis. For example if we only want to measure the complexity of the voter's strategy to successfully cast her vote, then we can abstract away from the cryptographic part of the model. On the other hand, when we want to take a closer look at the potential coercer and his capabilities some cryptographic details may be required. Also, it is impossible to predict in advance all the relevant aspects of the environment in which the protocol will be executed. This is even more the case when the protocol is embedded in a social environment, with human participants of unknown capabilities and heterogeneous motivations.
}

\begin{figure}[t]\centering
\includegraphics[scale=0.58]{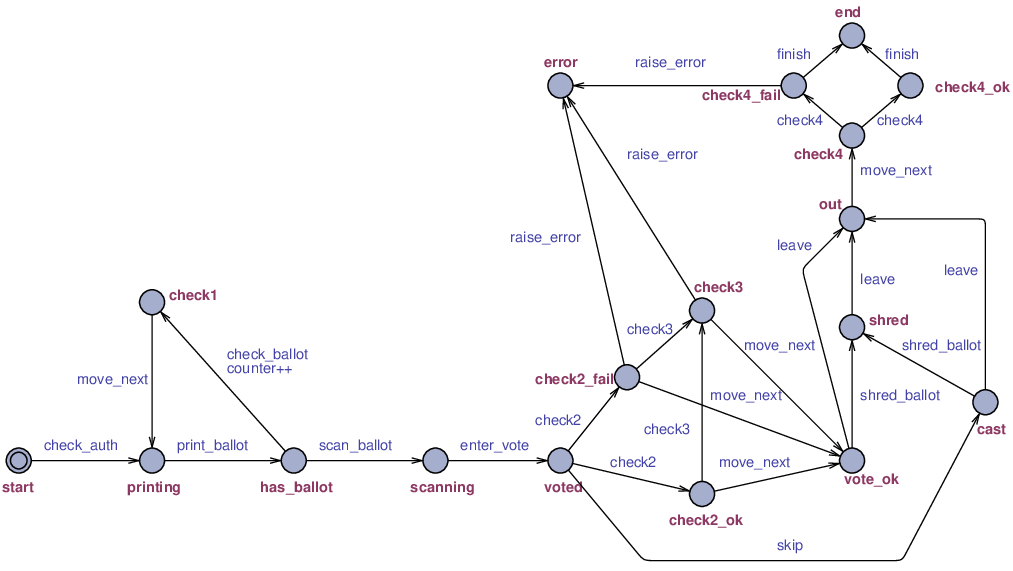}
\caption{\label{fig:votmodel} Voter model}
\end{figure}

\subsection{Modeling the Voting Process}

\extended{When the protocol is known, and the level of abstraction is decided, it is time to create a formal model of the voting procedure.}
The first step is to divide the description of the protocol into loosely coupled components, called agents. The partition is often straightforward: in our case, it will include the voter, the election infrastructure, the teller etc.
\extended{Depending on the level of abstraction abstraction that we chose, some models will be more detailed than others.}

For each agent we define its local model, which consists of locations (i.e., the local states of the agent) and labeled edges between locations (i.e., local transitions). A transition corresponds to an action performed by the agent. An example model of the voter can be seen in Figure~\ref{fig:votmodel}. When the voter has scanned her ballot and is in the state $scanning$ she can perform action $enter\_vote$, thus moving to the state $voted$. This model, as well as the others, has been created using the modeling interface of the \Uppaal model checker~\cite{Behrmann04uppaal-tutorial}. The locations in \Uppaal are graphically represented as circles, with initial locations marked by a double circle. The edges are annotated by colored labels: guards (green), synchronizations (teal) and updates (blue). The syntax of expressions is like that of C/C++. Guards enable the transition if and only if the guard condition evaluates to true. Synchronizations allow processes to synchronize over a common channel. Update expressions are evaluated when the transition is taken.

The global model of the whole system consists of a set of concurrent processes, i.e., local models of the agents.
The combination of the local models unfolds into a global model, where each global state represents a possible configuration of the local states of the agents.

\subsection{Natural Strategic Ability}

Many relevant properties of multi-agent systems refer to \emph{strategic abilities} of agents and their groups.
For example, voter-verifiability can be understood as the ability of the voter to check if her vote was registered and tallied correctly.
Similarly, receipt-freeness can be understood as the inability of the coercer, typically with help from the voter, to obtain evidence of how the voter has voted~\cite{Tabatabaei16expressing}.

Logics of strategic reasoning, such as \ATL and Strategy Logic, provide neat languages to express properties of agents' behavior and its dynamics, driven by individual and collective goals of the agents~\cite{Alur02ATL,Chatterjee10strategylogic,Mogavero14behavioral}.
For example, the \ATL formula $\coop{cust}\Sometm\prop{ticket}$
may be used to express that the customer $cust$ can ensure that he will eventually obtain a ticket, regardless of the actions of the other agents.
Semantically, the specification holds if $cust$ has a strategy whose every execution path satisfies $\prop{ticket}$ at some point in the future.
Strategies in a multi-agent system are understood as conditional plans, and play central role in reasoning about purposeful agents~\cite{Alur02ATL,Shoham09MAS}.
Formally, strategies are defined as \extended{functions from sequences of system states (i.e., possible histories of the game) to actions. A simpler notion of positional strategies, that we will use here, is defined by }functions from states to actions.
However, real-life processes often have millions or even billions of possible states, which allows for terribly complicated strategies -- and humans are notoriously bad at handling combinatorially complex objects.

To better model the way human agents strategize, the authors of~\cite{Jamroga19natstrat-aij,Jamroga19natstratii} proposed to use a more human-friendly representation of strategies, based on lists of condition-action rules.
The conditions are given by Boolean formulas\extended{ (for positional strategies) and regular expressions over Boolean formulas in the general case}.
Moreover, it was postulated that only those strategies should be considered whose complexity does not exceed a given bound.
This is consistent with classical approaches to commonsense reasoning~\cite{Davis15commonsense} and planning~\cite{Ghallab04planning}, as well as the empirical results on how humans learn and use concepts~\cite{Bourne70concepts,Feldman00conceptlearning}.
\nocite{Santos18SocialNormComplexity,Santos18phd}

\para{Natural strategies.}
Let $\Bool(\AP_a)$ be the set of Boolean formulas over atomic propositions $\AP_a$ observable by agent $a$.
In our case, $\AP_a$ consists of all the references to the local variables of agent $a$, as well as the global variables in the model.
We represent natural \extended{positional }strategies of agent $a$ by \emph{lists of guarded actions}, i.e., sequences of pairs $\gcmd{\phi_i}{\alpha_i}$ such that: (1) $\phi_i\in\Bool(\AP_a)$, and (2) $\alpha_i$ is an action available to agent $a$ in every state where $\phi_i$ holds.
Moreover, we assume that the last pair on the list is $\gcmd{\top}{\alpha}$ for some action $\alpha$, i.e., the last rule is guarded by a condition that will always be satisfied.
%
A \emph{collective natural strategy} for a group of agents $A=\set{a_1,\ldots,a_{|A|}}$ is a tuple of individual natural strategies $s_A = (s_{a_1}, \ldots, s_{a_{|A|}})$.
The set of such strategies is denoted by $\natstratsnorecl[A]$.

\extended{
  By $\size(s_a)$, we denote the number of guarded actions in $s_a$.
  Moreover, $cond_i(s_a)$ denotes the $i$th guard (condition) on the list, and $act_i(s_a)$ the corresponding action.
  Finally, $match(q,s_a)$ is the smallest $i\le \size(s_a)$ such that $q \models cond_i(s_a)$ and $act_i(s_a) \in d_a(q)$.
  That is, $match(q,s_a)$ matches state $q$ with the first condition in $s_a$ that holds in $q$, and action available in $q$.
}
The ``outcome'' function $out(q,s_A)$ returns the set of all paths (i.e., all maximal traces) that occur when coalition $A$ executes strategy $s_A$ from state $q$ onward, and the agents outside $A$ are free to act in an arbitrary way\short{.}\extended{:
  \begin{eqnarray*}
  	&& \mbox{}\hspace{-0.6cm} out(q,s_A) = \{\lambda=q_0q_1\dots \mid (q_0 = q)\ \land\ \forall_{i \geq 0} \exists_{\alpha_1,\dots,\alpha_{|\Agt|}}\ .\ \\
  	&& \mbox{}\hspace{-0.4cm}
  	(a\in A \Rightarrow \alpha_a = act_{match(q_i,s_a)}(s_a)) \land \\
  	&& \mbox{}\hspace{-0.4cm} (a\notin A \Rightarrow \alpha_a \in \mathit{available}_a(q_i))
  	\ \!\land\ \! (q_{i+1} = succ(q_i, \alpha_1,\!..., \alpha_{|\Agt|})) \}.
  \end{eqnarray*}
}

\para{Complexity of strategies.}
We will use the following complexity metric for strategies: $\cmplx{s_A} = \sum_{(\phi,\alpha)\in s_A} |\phi|$, with $|\phi|$ being the number of symbols in $\phi$, without parentheses.
That is, $\cmplx{s_A}$ simply counts the total length of guards in $s_A$.
Intuitively, the complexity of a strategy is understood as its level of sophistication. It corresponds to the mental effort needed to come up with the strategy, memorize it, and execute it.

\subsection{Specification of Properties Based on Natural Strategies}\label{sub:spec}

To reason about natural strategic ability, the logic $\natatl$ was introduced in~\cite{Jamroga19natstrat-aij,Jamroga19natstratii} with the following syntax:
\begin{center}
	$\varphi ::= \prop{p}  \mid  \lnot \varphi  \mid  \varphi\land\varphi
	\mid  \coop{A}^{\leq k}\Next\varphi  \mid  \coop{A}^{\leq k}\Sometm\varphi  \mid  \coop{A}^{\leq k}\Always\varphi  \mid  \coop{A}^{\leq k}\varphi\Until\varphi.$
\end{center}
where $A$ is a group of agents and $k \in \mathbb{N}$ is a complexity bound.
Intuitively, $\coop{A}^{\leq k}\gamma$ reads as ``coalition $A$ has a collective strategy of size less or equal than $k$ to enforce the property $\gamma$.''
The formulas of $\natatl$ make use of classical temporal operators: ``$\Next$'' (``in the next state''), ``$\Always$'' (``always from now on''), ``$\Sometm$'' (``now or sometime in the future''), and $\Until$ (strong ``until'').
For example, the formula $\coop{cust}^{\leq 10}\Sometm\prop{ticket}$ expresses that the customer can obtain a ticket by a strategy of complexity at most $10$. This seems more appropriate as a functionality requirement than to require the existence of \emph{any} function from states to actions.
The path quantifier ``for all paths''\extended{ from temporal logic} can be defined as $\Apath\gamma \equiv \coop{\emptyset}^{\leq 0}\gamma$.

We will use $\natatl$ to specify requirements on the voting system.
For example, voter-verifiability captures the ability of the voter to verify her vote after the election. In our case, this is represented by the $check4$ phase, hence we can specify voter-verifiability with the formula $\coop{voter}^{\leq k}\Sometm (\prop{check4\_ok} \vee \prop{error})$. The intuition is simple: the voter has a strategy of size at most $k$ to successfully perform $check4$ or else signal an error.
A careful reader can spot one problem with the formalization: it holds if the voter can signal an error regardless of the outcome of the check (and it shouldn't!).
A better specification is given by $\coop{voter}^{\leq k}\Sometm (\prop{check4\_ok} \vee \prop{check4\_fail})$.
Moreover, we can use the formula $\Apath\Always(\prop{check4\_fail} \then \coop{voter}^{\leq k}\Sometm\prop{error})$ to capture \extended{the property of }\emph{dispute resolution}.


The conceptual structure of receipt-freeness is different. In that case, we want to say that the voter has no way of proving how she has voted, and that the coercer (or a potential vote-buyer) does not have a strategy that allows him to learn the value of the vote. Crucially, this refers to the \emph{knowledge} of the coercer. To capture the requirement, we need to extend $\natatl$ with knowledge operators $K_a$, with $K_a\varphi$ expressing that agent $a$ knows that $\varphi$ holds. For instance, $K_{coerc} \prop{voted_i}$ says that the coercer knows that the voter has voted for the candidate $i$. Then, receipt-freeness can be formalized as
\\
\centerline{$\bigwedge_{i\in\mathit{Cand}}\lnot \coop{coerc,voter}^{\leq k} \Always (\prop{end} \rightarrow (K_{coerc} \prop{voted_i} \lor K_{coerc} \lnot \prop{voted_i}))$.}
\\
That means that the coercer and the voter have no strategy with complexity at most $k$ to learn, after the election is finished, whether the voter has voted for $i$ or not.
Note that this is only one of the possible formalization of the requirement.
For example, one may argue that, to violate receipt-freeness, it suffices that the coercer can detect \emph{whenever the voter has not obeyed}; he does not have to learn the exact value of her vote. This can be captured by the following formula: $\bigwedge_{i\in\mathit{Cand}}\lnot \coop{coerc,voter}^{\leq k} \Always ((\prop{end} \land \lnot\prop{voted_i}) \rightarrow K_{coerc} \lnot\prop{voted_i})$.

\subsection{Using Verification Tools to Facilitate Analysis}

The focus of this work is on modeling and specification; the formal analysis is done mainly by hand.
However, having the models specified in \Uppaal suggests that we can also benefit from its model checking functionality.
Unfortunately, the requirement specification language of \Uppaal is very limited, and allows for neither strategic operators nor knowledge modalities. Still, we can use it to verify concrete strategies if we carefully modify the input formula and the model. We will show how to do it in Section~\ref{sec:verification}.

\section{Use Case Scenario: \vVote}\label{sec:usecase}
Secure and verifiable voting is becoming more and more important for the democracy to function correctly.
\extended{
  Voting protocols are often tested in small-scale elections or in very special scenarios. This allows the officials to monitor how people react, and the scientists to conduct an analysis of the election afterwards.}
In this paper, we analyze the \vVote implementation of \Pret which was used for remote voting and voting of handicapped persons in the Victorian elections in November 2014~\cite{Culnane15vvote}.
The main idea of the \Pret protocol focuses on encoding the vote using a randomized candidate list. In this protocol the ballot consists of two parts: the randomized order of candidates (left part) and the list of empty checkboxes along with the number encoding the order of the candidates (right part). The voter cast her vote in an usual way, by placing a cross in the right hand column against the candidate of her choice. After that she tears the ballot in two parts, destroys the left part, cast the right one and takes the copy of it as her receipt. After the election her vote appears on the Web Bulletin Board as the pair of the encoding number and the marked box, which can be compared with the receipt for verification.
We look at the whole process, from the voter entering the polling station, to the verification of her vote on the public Web Bulletin Board (WBB).

After entering the polling station, the Poll Worker (PW) authenticates the voter (using the method prescribed by the appropriate regulations), and sends a print request to the Print On Demand (PON) device specifying the district/region of the voter. If the authentication is valid (state $printing$) then the PON retrieves and prints an appropriate ballot for the voter, including a Serial Number (SN) and the district, with a signature from the Private Web Bulletin Board (PWBB). The PWBB is a robust secure database which receives messages, performs basic validity checks, and returns signatures.
After that, the voter may choose to check and confirm the ballot. This involves demanding a proof that the ballot is properly formed, i.e., that the permuted candidate list corresponds correctly to the cipher-texts on the public WBB for that serial number. The WBB is an authenticated public broadcast channel with memory. If the ballot has a confirmation check, the voter returns to the printing step for a new ballot (transition from state $check1$ to $printing$).

Having obtained and possibly checked her ballot (state $has\_ballot$), the voter can scan it by showing the ballot bar code to the Electronic Ballot Marker (EBM). Then, she enters her vote (state $scanning$) via the EBM interface.
\extended{The EBM is a computer that assists the user in filling in a \Pret ballot.}
The EBM prints on a separate sheet the voter’s receipt with the following information:
(i) the electoral district,
(ii) the Serial Number,
(iii) the voter's vote permuted appropriately to match the \Pret ballot,
and (iv) a QR code with this data and the PWBB signature.

Further, the voter must check the printed vote against the printed candidate list. In particular, she checks that the district is correct and the Serial Number matches the one on the ballot form. If all is well done, she can optionally check the PWBB signature, which covers only the data visible to the voter. 
Note that, if either $check2$ or $check3$ fails, the vote is canceled using the cancellation protocol. If everything is OK, the voter validates the vote, shreds the candidate list, and leaves the polling station. Finally, the voter can check her vote on the WBB after the election closes. She only needs to check the SR and the order of her preference numbers.



\section{Models}\label{sec:models}
In this section we present models of a simplified version of \vVote, focusing on the steps that are important from the voter's perspective.
We use \Uppaal as the modeling tool because of its flexible modeling language and user-friendly GUI.

\short{
	\begin{figure}[t]\centering
  		\includegraphics[scale=0.7]{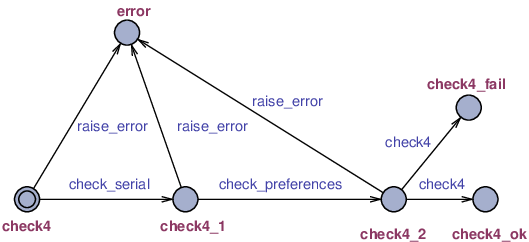}
			\caption{\label{fig:votcheck4} Voter refinement: phase check4}
  \end{figure}
}

\subsection{Voter Model}\label{models:voter}

The model already presented in Figure~\ref{fig:votmodel} captures the voter actions from entering the polling station to casting her vote, going back home and verifying her receipt on the web bulletin board. As shown in the model, some actions, i.e. additional checks, are optional for the voter. Furthermore, to simulate the human behavior we added some additional actions, not described in the protocol itself. For example the voter can skip even obligatory steps, like $check2$. This is especially important, as $check2$ may be the most time-consuming action for the voter and many voters may skip it in the real life.
\extended{
  To further simulate the real-life behavior of the voters, for each state we added a loop action labeled as $wait$, to allow the voter to wait in any state as long as she wants. We omit this loops from the model picture for the clarity of the presentation.}
After every check, the voter can signal an error, thus ending up in the $error$ state. The state represents communication with the election authority, signaling that the voter could not cast her vote or a machine malfunction was detected.

\subsection{Refinements of the Voter Model}

The model shown in Figure~\ref{fig:votmodel} is relatively abstract.
For example, $check4$ is shown as an atomic action, but in fact it requires that the voter compares data from the receipt and the WBB. In order to properly measure the complexity of the voter strategies, it is crucial to consider different levels of granularity.

\extended{
  \begin{figure}[t]
	\centering
  	\includegraphics[scale=1.0]{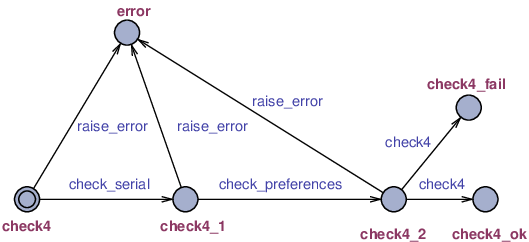}
  	\caption{\label{fig:votcheck4} Voter refinement: phase check4 (verification)}
  \end{figure}
}

\short{
  \begin{figure}[t]
  \hspace{-0.7cm}
  \begin{minipage}[b]{6cm}
  	\centering
  	\includegraphics[scale=0.7]{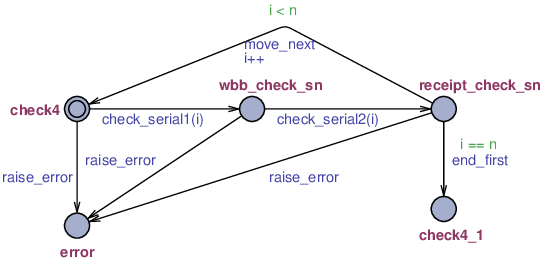}
  	\caption{\label{fig:check4sn} Serial number check}
  \end{minipage}
  \hspace{1cm}
  \begin{minipage}[b]{6cm}
  	\centering
  	\includegraphics[scale=0.7]{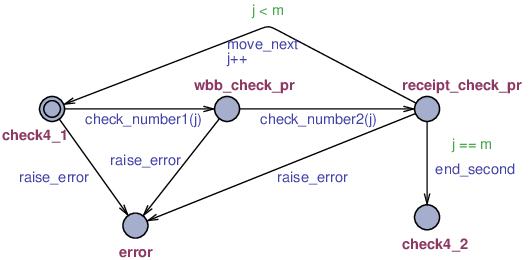}
  		\caption{\label{fig:check4pp} Preferences order check}
  \end{minipage}
  \end{figure}
}

\para{Check4 phase.}
Recall that this is the last phase in the protocol and it is optional.
Here, the voter can check that the printed receipt matches her intended vote on the WBB. This includes checking that the serial numbers match (action $check\_serial$), and that the printed preferences order match the one displayed on the WBB (action $check\_preferences$). So, if both steps succeed, then the voter reach state $check4\_ok$. The whole model for this phase is presented in Figure~\ref{fig:votcheck4}. \short{Other phases, like $check2$ can be presented in a similar way.}
	
\extended{
  \begin{figure}[t]
  	\centering
  	\includegraphics[scale=1.0]{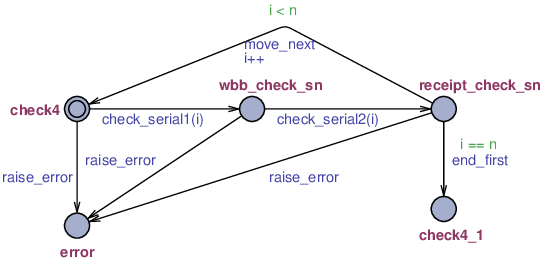}
  	\caption{\label{fig:check4sn} Voter refinement: serial number check. Label $check\_serial(i)$ depicts checking the $i$th symbol of the serial number on the receipt and WBB}
  \end{figure}
}

\para{Serial number phase.}
In some cases the model shown in Figure~\ref{fig:votcheck4} may still be too general. Depending on the length of the serial number we can have different levels of difficulty. For example comparing two alphanumeric sequences of length 2 is easier then comparing such sequences of length 10. To express this concept, we split this step into atomic actions: $check\_serial1(i)$ for checking the $i$th symbol on the WBB, and $check\_serial2(i)$ for checking the $i$th symbol on the receipt. The resulting model is shown in Figure~\ref{fig:check4sn}, where $n$ is the length of the serial number.

\para{Preferences order phase.}
Similarly to comparing the two serial numbers, verifying the printed preferences can also be troublesome for the voter. In order to make sure that her receipt matches the entry on the WBB, the voter must check each number showing her preference. Actions $check\_number1(i)$ and $check\_number2(i)$ mean checking number on the WBB and on the receipt, respectively. This is shown on the model in Figure~\ref{fig:check4pp}, where $m$ is the number of candidates in the ballot.

\extended{
  \begin{figure}[t]
  	\begin{center}
  		\includegraphics[scale=1.0]{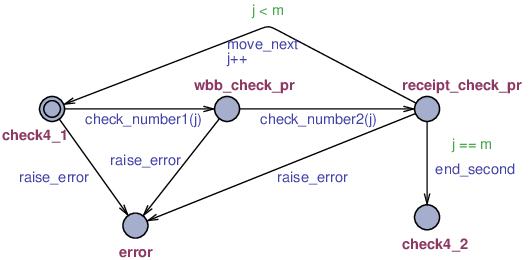}
  		\caption{\label{fig:check4pp} Voter refinement: preferences order check. Label $check\_number(i)$ means checking the $i$th position on the receipt and WBB}
  	\end{center}
  \end{figure}

  \para{Preferences order phase.}
  Similarly to comparing the two serial numbers, verifying the printed preferences can also be troublesome for the voter. In order to make sure that her receipt matches the entry on the WBB, the voter must check each number showing her preference. Actions $check\_number1(i)$ and $check\_number2(i)$ mean checking number on the WBB and on the receipt, respectively. This is shown on the model in Figure~\ref{fig:check4pp}, where $m$ is the number of candidates in the ballot.

  	\begin{figure}[t]
  		\begin{center}
  					\includegraphics[scale=1.0]{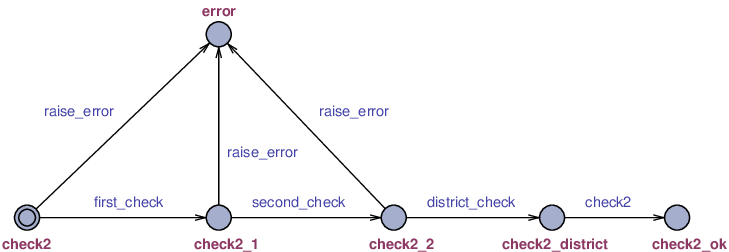}
  			\caption{\label{fig:votcheck2} Detail on the voter model: phase check 2.}
  		\end{center}
  	\end{figure}
  	
  \para{Check2 phase.}
  We now focus our attention on the $check2$ phase. Recall that this is the only one phase that the voter has to check, all the others are optional.
  	Here, the voter should check that the printed receipt matches his intended vote. This includes checking that the serial numbers match (action $first check$), and that the printed preferences match his intended vote arranged according to the candidate order on his ballot (action $second check$). So, if both the steps are exceeded, then the voter checks that the district is correct. A refined model for this phase is shown in Figure~\ref{fig:votcheck2}.
}

\begin{figure}[t]
\begin{minipage}[b]{4cm}
	\centering
	\includegraphics[scale=1.0]{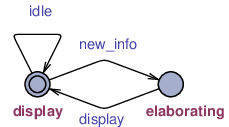}
	\caption{\label{fig:publicWBB} Public WBB}
\end{minipage}
\begin{minipage}[b]{4cm}
	\centering
	\includegraphics[scale=1.0]{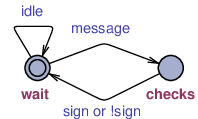}
	\caption{\label{fig:privateWBB} Private WBB}
\end{minipage}
	\begin{minipage}[b]{4cm}
	\centering
	\includegraphics[scale=1.0]{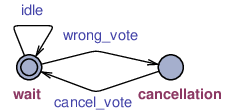}
	\caption{\label{fig:cancelstation} Cancel station}
\end{minipage}
\end{figure}

\subsection{Voting Infrastructure}\label{models:system}

The voter is not the only entity taking part in the election procedure. The election infrastructure and the electronic devices associated with it constitute a significant part of the procedure. Since there are several components involved in the voting process, we decided to model each component as a separate agent.
\short{The models of the Public WBB, Private WBB, the cancel station, the print-on-demand printer, and the EBM are shown in Figures~\ref{fig:publicWBB}--\ref{fig:ebm}.}

\extended{
  \para{Public WBB.}
  In Figure~\ref{fig:publicWBB} we present the public WBB. Simply, this component displays a new information when it receives a new one.
  		
  \para{Private WBB.}
  In Figure~\ref{fig:privateWBB} we present the private WBB. Here, for each message received, the system component decides to sign or not the message.

  \para{Cancel station.}	
  In Figure~\ref{fig:cancelstation} we present the cancel station. This component is a supervised interface for canceling a vote that has not been properly submitted or has not received a valid PWBB signature.
  	
  The three models described above have a very similar structure. In fact, each component waits for an event outside it and performs an action that returns in all the cases in the initial state.
  	
  \para{Print-on-demand printer.}
  In Figure~\ref{fig:print} we show the steps of the printer. Starting from the state $wait$ in which the printer does always idle, the situation changes when another agent sends a print request (an event). In the state $start$ the printer checks whether the agent that has made the request has an account. If this is the case the printer prints the document, passes in the end state and then returns in wait. Otherwise, the printer does not accept the request and returns in $wait$.
  	
  \para{EBM.}
  Finally, in Figure \ref{fig:ebm} we present the interactions regarding the EBM. First, the EBM receives a ballot (event) and controls if the bar code is correct. If the control is exceeded, the EBM checks whether the vote is OK and then prints it. In all the other wrong cases the EBM passes in the $error$ state and then returns in the initial state ($wait$).
}

\extended{
\begin{figure}[t]
	\begin{center}
				\includegraphics[scale=1.0]{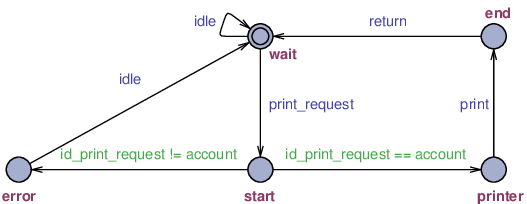}
		\caption{\label{fig:print} Print-on-demand printer}
	\end{center}
\end{figure}

\begin{figure}[t]
	\begin{center}
				\includegraphics[scale=1.0]{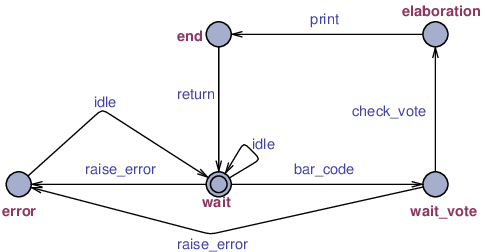}
		\caption{\label{fig:ebm} Electronic Ballot Marker}
	\end{center}
\end{figure}
}

\short{
  \begin{figure}[t]
    \begin{minipage}[b]{6cm}
    	\centering
    	\includegraphics[scale=0.7]{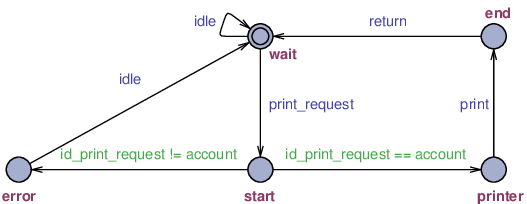}
    	\caption{\label{fig:print} Print-on-demand printer}
    \end{minipage}
    \hspace{0.5cm}
    \begin{minipage}[b]{6cm}
    	\centering
    	\includegraphics[scale=0.7]{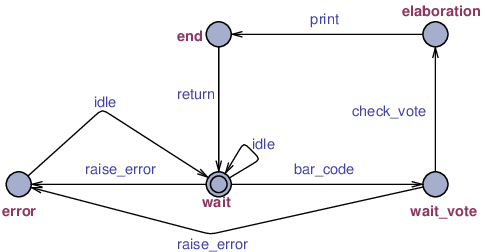}
    	\caption{\label{fig:ebm} Electronic Ballot Marker (EBM)}
    \end{minipage}
   \end{figure}
}

\subsection{Coercer Model}\label{models:coercer}

To model the coercer, we first need to determine his exact capabilities. Is he able to interact with the voter, or only with the system? Should he have full control over the network, like the Dolev-Yao attacker, or do we want the agent to represent implicit coercion, where the relatives or subordinates are forced to vote for the specified candidate.
There are many possibilities, and to this end we have decided that one model for the \Coercer agent is not enough.
Because of that and due to lack of space, we omit the details, and only describe the possible actions of the coercer:
\begin{itemize}
	\item[$\bullet$] $Coerce(v,ca)$: the coercer coerces the voter $v$ to vote for his candidate $ca$;
	\item[$\bullet$] $ModifyBallot(v,ca)$: the coercer modifies the ballot for $v$ by setting $ca$;
	\item[$\bullet$] $RequestVote(v)$: the coercer requests a vote from $v$;
	\item[$\bullet$] $Punish(v)$: the coercer punishes $v$;
	\item[$\bullet$] $Infect$: the coercer infects the voting machine with malicious code;
	\item[$\bullet$] $Listen(v)$: the coercer listens to the vote of $v$ from the voting machine;
	\item[$\bullet$] $Replace(v,ca)$: the coercer replaces the vote of $v$ with $ca$.
\end{itemize}

Some of the actions may depend on each other. For example, $Listen$ and $Replace$ actions should be executed only after the $Infect$ action has succeed, as the \Coercer needs some kind of access to the voting machine.

	
	
	
	

\section{Strategies and Their Complexity}\label{sec:strats}

There are many possible objectives for the participants of a voting procedure. A voter's goal could be to just cast her vote, another one could be to make sure that her vote was correctly counted, and yet another one to verify the election results. The same goes for the coercer: he may just want to make his family vote in the ``correct,'' or to change the outcome of the election. In order to define different objectives, we can use formulas of \natatl and look for appropriate natural strategies, as described in Section~\ref{sec:methodology}.
More precisely, we can fix a subset of the participants and their objective with a formula of \natatl, find the smallest strategy that achieves the objective, and compute its size. The size of the strategy will be an indication of how hard it is to make sure that the objective is achieved.

An example of specification that the voter wants to achieve is the verifiability of her vote. Given the model in Figure \ref{fig:votmodel}, we can use the formula $\coop{v}^{\leq k} \F check4\_ok$ to check whether the voter has a natural strategy of size less or equal than $k$ to verify that her receipt has the same information displayed in the public WBB. If we want to check only if the voter has a natural strategy to verify her receipt, i.e. we want also consider the case in which the voter's receipt and the information in the public WBB are different, we can consider the formula  $\coop{v}^{\leq k} \F (check4\_ok \vee check4\_fail)$ as we discussed in Section \ref{sub:spec}.

Note that it is essential to fix the granularity level of the modeling right. When shifting the level of abstraction, we obtain significantly different ``measurements'' of strategic complexity. This is why we proposed several variants of the voter model in Section~\ref{sec:models}. In this section, we will show how it affects the outcome of the analysis.
	
In the following we take another look at the previously defined models and try to list possible strategies for the participants.

\subsection{Strategies for the Voter}\label{sec:voter-strats}
	
	In this section we focus on natural strategies for voter-verifiability.
	Given the model in Figure~\ref{fig:votmodel}, we can analyze an example of natural strategy that can be used by the voter to achieve the end of the voting procedure, i.e., to satisfy the NatATL formula $\varphi_1 = \coop{v}^{\leq k} \F \prop{end}$. Clearly, $\varphi_1$ specifies the fact that the voter has a natural strategy of size less or equal than $k$ (captured by $\coop{v}^{\leq k}$) to reach sooner or later (captured by the eventually operator $\F$) the end of procedure (i.e. the state labeled with atom $\prop{end}$).

	\begin{NatStr}\label{mainstrat}
		A strategy for the voter is:
		\begin{enumerate}
			\item $\prop{has\_ballot} \thus scan\_ballot$
			\item $\prop{scanning} \thus enter\_vote$
			\item $\prop{voted} \thus check2$
			\item $\prop{check2\_ok} \vee \prop{check2\_fail} \vee \prop{out} \thus move\_next$
			\item $\prop{vote\_ok} \thus shred\_ballot$
			\item $\prop{shred} \thus leave$
			\item $\prop{check4} \thus check4$
			\item $\prop{check4\_ok} \vee \prop{check4\_fail} \thus finish$
			\item $\top \thus \star$
		\end{enumerate}
	\end{NatStr}

Recall that the above is an ordered sequence of guarded commands. The first condition (guard) that evaluates to \emph{true} determines the action of the voter. Thus, if the voter has the ballot and she has not scanned it (proposition $\prop{has\_ballot}$), she scans the ballot. If $\prop{has\_ballot}$ is false and $\prop{scanning}$ is true then she enters her vote, and so on.
If all the preconditions except $\top$ are false, then she executes an arbitrary available action (represented by the wildcard $\star$). For example, the voter will do $print\_ballot$ at the state $printing$, where the voter needs to wait while the Pool Walker identifies her and generates a new ballot.

	In Natural Strategy \ref{mainstrat}, we have $9$ guarded commands in which the command (4) costs $5$ since in its condition there are five symbols (three atoms plus two disjunctions), the command (7) costs $3$ since in its condition there are three symbols (two atoms plus the disjunction), while the other guarded commands cost $1$, so the total complexity is $1 \cdot 7 + 3 \cdot 1 + 5 \cdot 1 = 15$. So, the formula $\varphi_1$ is true with any $k$ of $15$ or more. Further, by Natural Strategy \ref{mainstrat}, starting from the state $\prop{has\_ballot}$, the voter needs of $9$ steps to achieve the state $\prop{end}$.
	
	Note that Natural Strategy \ref{mainstrat} can be also used to demonstrate that the formula $\psi = \coop{v}^{\leq k} \F (\allowbreak\prop{check4\_ok} \vee \prop{check4\_fail})$ holds. In that case, we can reduce the size of the strategy by removing the guarded command (8). Thus, $\psi$ is satisfied even for $k \geq 12$.
	
	For completeness, in what follows, we show a natural strategy with the additional guarded commands in case the voter wants to do the optional phases $check1$ and $check3$, i.e., we want to satisfy the formula $\varphi_2 = \coop{v}^{\leq k} \F (\prop{checked1} \land \prop{checked3} \land \prop{end})$. In particular, $\varphi_2$ checks whether there exist a natural strategy for the voter such that sooner or later she does $check1$, $check3$, and ends the whole voting process. Note that, apart from the standard propositions like $check1$, we also add their persistent version like $checked1$, i.e., once it gets true, it remains always true.
		
		\begin{NatStr}\label{mainstratwithchecks}
		A strategy for the voter that considers the optional phases $check1$ and $check3$ is:
		\begin{enumerate}
			\item $\prop{has\_ballot} \land \prop{counter == 0} \thus check1$
			\item $\prop{has\_ballot} \thus scan\_ballot$
			\item $\prop{scanning} \thus enter\_vote$
			\item $\prop{voted} \thus check2$
			\item $\prop{check2\_ok} \vee \prop{check2\_fail} \thus check3$
			\item $\prop{check1} \vee \prop{check3} \vee \prop{out} \thus move\_next$
			\item $\prop{vote\_ok} \thus shred\_ballot$
			\item $\prop{shred} \thus leave$
			\item $\prop{check4} \thus check4$
			\item $\prop{check4\_ok} \vee \prop{check4\_fail} \thus finish$
			\item $\top \thus \star$
		\end{enumerate}
	\end{NatStr}


	In Natural Strategy \ref{mainstratwithchecks}, we introduce the verification of $check1$ and $check3$. To do this we add two new guarded commands (5) and (6), and we update (1) by adding a control on a counter to determine if $check1$ is done or not. Here, we have a total complexity of $1 \cdot 7 + 3 \cdot 3 + 5 \cdot 1 = 21$. So, the formula $\varphi_2$ is true for any $k \geq 21$. Further, by Natural Strategy \ref{mainstratwithchecks}, starting from the state $\prop{has\_ballot}$, the voter needs of $13$ steps to achieve the state $\prop{end}$.
	
	An important aspect to evaluate in this subject concerns the detailed analysis of $check4$. Some interesting questions on this analysis could be: how does the voter perform $check4$? How does she compare the printed preferences with the information in the public WWB? These questions open up several scenarios both from a strategic point of view and from the model to be used. From a strategic point of view, we could consider a refinement of Natural Strategy \ref{mainstrat}, in which the action $check4$ is evaluated as something of atomic. If we consider that the $check4$ includes: comparing preferences with the information in the public WWB and checking the serial number, we can already divide the single action into two different actions for each of the checks to be performed.
	So, given the model in Figure~\ref{fig:votcheck4}, we can refine the natural strategy for the voter, as follows.
	
	\begin{NatStr}\label{mainstratcheck4}
		A strategy for the voter that refines $check4$ is:
		\begin{enumerate}
			\item $\prop{has\_ballot} \thus scan\_ballot$
			\item $\prop{scanning} \thus enter\_vote$
			\item $\prop{voted} \thus check2$
			\item $\prop{check2\_ok} \vee \prop{check2\_fail} \vee \prop{out} \thus move\_next$
			\item $\prop{vote\_ok} \thus shred\_ballot$
			\item $\prop{shred} \thus leave$
			\item $\prop{check4} \thus check\_serial$
			\item $\prop{check4\_1} \thus check\_preferences$
			\item $\prop{check4\_2} \thus check4$
			\item $\prop{check4\_ok} \vee \prop{check4\_fail} \thus finish$
			\item $\top \thus \star$
		\end{enumerate}
	\end{NatStr}
	
	In Natural Strategy \ref{mainstratcheck4}, we have $11$ guarded commands in which all the conditions are defined with a single atom but (4) in which there is a disjunction of three atoms and (10) in which there is a disjunction of two atoms. So, the total complexity is $1 \cdot 9 + 3 \cdot 1 + 5 \cdot 1 = 17$.
	
	To verify that the voter does each step of $check4$, we need to provide a formula that verifies atoms $\prop{check4}$, $\prop{check4\_1}$, and $\prop{check4\_2}$. To do this in NatATL, we use the formula $\varphi_3 = \coop{v}^{\leq k} \F (\prop{checked4} \land \prop{checked4\_1} \land \prop{checked4\_2})$.
	Note that $\varphi_3$ is true for any $k \geq 17$; one can use Natural Strategy \ref{mainstratcheck4} to demonstrate that.

	In addition, to increase the level of detail, one could consider that the voter checks the preferences and the serial number one by one in an ordered fashion. So, given the models in Figures~\ref{fig:check4sn} and \ref{fig:check4pp}, we can consider a formula that checks whether the voter has a strategy that satisfies the following properties:
	
	\begin{enumerate}
		\item sooner or later she enters in the $check4$ phase;
		\item she verifies a symbols of the SN in the public WBB and in her receipt;
		\item she does (2) until the last symbol of the serial number is verified;
		\item she does a similar approach as in (2)-(3) for the verification of preferences;
		\item she finishes the whole procedure.
	\end{enumerate}

	This can be captured by the formula
	$\varphi_4 = \coop{v}^{\leq k} \F (\prop{checked4} \land \prop{wbb\_checked\_sn} \land \prop{receipt\_checked\_sn} \land \prop{checked4\_1} \land \prop{wbb\_checked\_pr} \land \prop{receipt\_checked\_pr} \land \prop{checked4\_2})$. So, we can define a natural strategy that satisfies $\varphi_4$, as follows.
	
	\begin{NatStr}\label{mainstratcheck4ref}
		A strategy for the voter that still refines $check4$ is:
		\begin{enumerate}
			\item $\prop{has\_ballot} \thus scan\_ballot$
			\item $\prop{scanning} \thus enter\_vote$
			\item $\prop{voted} \thus check2$
			\item $\prop{vote\_ok} \thus shred\_ballot$
			\item $\prop{shred} \thus leave$
			\item $\prop{out} \vee \prop{check2\_ok} \vee \prop{check2\_fail} \vee \prop{receipt\_check\_sn} \vee \prop{receipt\_check\_pr} \thus move\_next$
			\item $\prop{check4} \thus check\_serial1$
			\item $\prop{wbb\_check\_sn} \thus check\_serial2$
			\item $\prop{receipt\_check\_sn} \land \prop{i == n} \thus end\_first$
			\item $\prop{check4\_1} \thus check\_number1$
			\item $\prop{wbb\_check\_pr} \thus check\_number2$
			\item $\prop{receipt\_check\_pr} \land \prop{j == m} \thus end\_second$
			\item $\prop{check4\_2} \thus check4$
			\item $\prop{check4\_ok} \vee \prop{check4\_fail}\thus finish$
			\item $\top \thus \star$
		\end{enumerate}
	\end{NatStr}
	
	To conclude,
	the above natural strategy has $15$ guarded commands in which the conditions in (9), (12), and (14) are conjunctions of two atoms, the condition in (6) is a disjunction of five atoms, and all the other conditions are defined with a single atom. Therefore, the complexity of Natural Strategy \ref{mainstratcheck4ref} is $1 \cdot 11 + 3 \cdot 3 + 9 \cdot 1 = 29$. So, the formula $\varphi_4$ is true with $k \geq 29$.

\subsection{Counting Other Kinds of Resources}

So far, we have measured the effort of the voter by how complex strategies she must execute. This helps to estimate the mental difficulty related, e.g., to voter-verifiability. However, this is not the only source of effort that the voter has to invest. Verifying one's vote might require money (for example, if the voter needs to buy special software or a dedicated device), computational power, and, most of all, time.
Here, we briefly concentrate on the latter factor.

For a voter's task expressed by the \natatl formula $\coop{v}^{\le k}\Sometm\varphi$ and a natural strategy $s_v$ for the voter, we can estimate the time spent on the task by the number of transitions necessary to reach $\varphi$. That is, we take all the paths in $out(q,s_v)$, where $q$ is the initial state of the procedure. On each path, $\varphi$ must occur at some point. We look for the path where the first occurrence of $\varphi$ happens \emph{latest}, and count the number of steps to $\varphi$ on that path.
We will demonstrate how it works on the last two strategies from Section~\ref{sec:voter-strats}.

For Natural Strategy \ref{mainstratcheck4}, starting from the starting state, the voter needs of $9 + 2 = 11$ steps to achieve $\prop{check4} \land \prop{wbb\_check\_sn} \land \prop{receipt\_check\_sn} \land \prop{check4\_1} \land \prop{wbb\_check\_pr} \land \prop{receipt\_check\_pr} \land \prop{check4\_2}$. More precisely, $9$ steps are needed to achieve $check4$ in the local model shown in Figure~\ref{fig:votmodel}, and $2$ more steps to reach $check4\_2$ in the refinement of the final section of the procedure (see Figure~\ref{fig:votcheck4}).

For Natural Strategy~\ref{mainstratcheck4ref}, the calculation is slightly more complicated. Starting from state $\prop{start}$, the voter needs of $9 + ((2 \cdot n) + 1) + ((2 \cdot m) + 1)$ steps to achieve her goals, where $n$ and $m$ are the sizes of the serial number and the list of preferences, respectively. In particular, $9$ steps are needed to achieve $check4$ in Figure~\ref{fig:votmodel}, then $((2 \cdot n) + 1)$ steps to achieve $check4\_1$ in Figure~\ref{fig:check4sn}, and finally $((2 \cdot m) + 1)$ steps to achieve $check4\_2$ in Figure~\ref{fig:check4pp}.

\subsection{Strategies for the Coercer}

For the coercer, we can start the analysis by considering the basic setting in which he requests the vote and if the voter does not give him the ballot or the vote is different with the one imposed by him, then he punish the voter. This reasoning is captured by the following natural strategy.
	
	\begin{enumerate}
		\item $\neg \prop{coerced_{v}} \thus Coerce(v,ca)$
		\item $\prop{coerced_{v}} \land \neg \prop{requested_v} \thus RequestVote(v)$
		\item $\prop{coerced_{v}} \land \prop{requested_v} \land \neg \prop{punished_v} \land (\prop{ca_v} \neq \prop{ca} \lor \prop{not\_show_v}) \thus Punish(v)$
	\end{enumerate}

	The total complexity of the above natural strategy is $16$ since (1) has $2$ symbols (atom + negation), (2) has $4$ symbols (two atoms + conjunction + negation), and (3) has $10$ symbols (five atoms + three conjunctions + negation + disjunction).
		
	Another intervention of the coercer regards the possibility to infect one or more machines and to replace the vote of a voter.
	
	\begin{enumerate}
		\item $\neg \prop{infected} \thus Infect$
		\item $\prop{infected} \land \neg \prop{replaced_{v}} \thus Replace(v,ca)$
	\end{enumerate}

	In this case the complexity is $6$ since we have three atoms involved, two negations, and a conjunction.
	
	We can extend the above by considering that the coercer can infect the machine and then can coerce the voter. So, if the coercer see a different vote on the machine he can punish the voter. Note that this setting is interesting in the case the coercer has infected a machine that only displays informations without having the power to modify the vote. 
	
	\begin{enumerate}
		\item $\neg \prop{infected} \thus Infect$
		\item $\neg \prop{coerced_{v}} \thus Coerce(v,ca)$
		\item $\prop{infected} \land \prop{listen_{v}} \neq \prop{ca} \thus Punish(v)$
	\end{enumerate}

	In the above, the complexity is $7$, where (1) and (2) have complexity $2$ and (3) has complexity $3$.

\section{Automated Verification of Strategies}\label{sec:verification}
In this section we explain how the model checking functionality of \Uppaal can be used for an automated verification of the strategies presented in Section~\ref{sec:strats}. To verify selected formulas and the corresponding natural strategies, we need to modify several things: \emph{(i)} the formula, \emph{(ii)} the natural strategy and finally, \emph{(iii)} the model. We start by explaining the required modifications step by step.

\para{Formula.} To specify the required properties for the protocol, we have used a variant of strategic logic, as it is one of the most suited logic to specify properties for agents in an intuitive way. However, \Uppaal does not support \natatl, so the formula needs to be modified accordingly. In the formula we replace the strategic operator $\coop{A}^{\leq k}$ with an universal path quantifier A. For example, we can consider the formula $\varphi_1$ used in Section \ref{sec:strats}. In this context, we produce the CTL formula $\varphi'_1 = \A \F \prop{end}$.

\para{Natural Strategy.} The natural strategy needs to be modified so that all the guard conditions are mutually exclusive.
To this end, we go through the preconditions from top to bottom, and refine them by adding the negated preconditions from all the previous guarded commands. 
For example, Natural Strategy~\ref{mainstrat} is modified as follows:
		\begin{enumerate}
			\item $\prop{has\_ballot} \thus scan\_ballot$
			\item $\lnot \prop{has\_ballot} \land \prop{scanning} \thus enter\_vote$
			\item $\lnot \prop{has\_ballot} \land \lnot \prop{scanning} \land \prop{voted} \thus check2$
			\item $\lnot \prop{has\_ballot} \land \lnot \prop{scanning} \land \lnot \prop{voted} \land (\prop{check2\_ok} \vee \prop{check2\_fail} \vee \prop{out}) \thus move\_next$
			\item $\lnot \prop{has\_ballot} \land \lnot \prop{scanning} \land \lnot \prop{voted} \land \lnot (\prop{check2\_ok} \vee \prop{check2\_fail} \vee \prop{out}) \land \prop{vote\_ok} \thus shred\_ballot$
			\item $\lnot \prop{has\_ballot} \land \lnot \prop{scanning} \land \lnot \prop{voted} \land \lnot (\prop{check2\_ok} \vee \prop{check2\_fail} \vee \prop{out}) \land \lnot \prop{vote\_ok} \land \prop{shred} \thus leave$
			\item $\lnot \prop{has\_ballot} \land \lnot \prop{scanning} \land \lnot \prop{voted} \land \lnot (\prop{check2\_ok} \vee \prop{check2\_fail} \vee \prop{out}) \land \lnot \prop{vote\_ok} \land \lnot \prop{shred} \land (\prop{check4\_ok} \vee \prop{check4\_fail}) \thus finish$
			\item $\top \thus \star$
		\end{enumerate}

\para{Model.} To verify the selected strategy, we fix it in the model by adding the preconditions of the guards to the preconditions of the corresponding local transitions in the voter's model. Thus, we effectively remove all transitions that are not in accordance with the strategy.
This way, only the paths that are consistent with the strategy will be considered by the model-checker. 

\para{Levels of granularity.}
As we showed in Section \ref{sec:models}, it is often important to have variants of the model for different levels of abstraction. To handle those in \Uppaal, we have used synchronizations edges. For example, to have a more detailed version of the phase $check4$, we added synchronization edges in the voter model (Figure \ref{fig:votmodel}) and in the $check4$ model (Figure \ref{fig:votcheck4}). Then, when going through the $check4$ phase in the voter model, \Uppaal will proceed to the more detailed model and come back after getting to its final state.

\para{Running the verification.}
We have modified the models, formulas, and strategies from Sections~\ref{sec:models} and \ref{sec:strats} following the above steps. Then, we used \Uppaal to verify that Natural Strategies~1--4 indeed enforce the prescribed properties. The tool reported that each formula holds in the corresponding model. The execution time was always at most a few seconds.

\section{Conclusions}\label{sec:conclusions}
In the analysis of a voting protocols it is important to make sure that the voter has a strategy to use the functionality of the protocol. 
That is, she has a strategy to fill in and cast her ballot, verify her vote on the bulletin board, etc.
However, this is not enough: it is also essential to see how hard that strategy is. 
In this paper, we propose a methodology that can be used to this end. 
One can assume a natural representation of the voter's strategy, and to measure its complexity as the size of the representation.

We mainly focus on one aspect of the voter's effort, namely the mental effort needed to produce, memorize, and execute the required actions. 
We also indicate that there are other important factors, like the time needed to execute the strategy or the financial cost of the strategy. 
This may lead to tradeoffs where optimizing the costs with respect to one resource leads to higher costs in terms of another resource. 
Moreover, resources can vary in their importance for different agents. For example, time may be more important for the voter, while money more relevant when we analyze the strategy of the coercer.
We leave a closer study of such tradeoffs for future work.

Another interesting extension would be to further analyze the parts of the protocol where the voter compares two numbers, tables, etc. As the voter is a human being, it is natural for her to make a mistake. Furthermore, the probability of making a mistake at each step can be added to the model to analyze the overall probability of successfully comparing two data sets by the voter.

Finally, we point out that the methodology proposed in this paper can be applied outside the e-voting domain. 
For example, one can use it to study the usability of policies for social distancing in the current epidemic situation, and whether they are likely to obtain the expected results.


\bibliographystyle{plain}
\bibliography{wojtek,wojtek-own}

\begin{thebibliography}{10}

\bibitem{Alur02ATL}
R.~Alur, T.~A. Henzinger, and O.~Kupferman.
\newblock {A}lternating-time {T}emporal {L}ogic.
\newblock {\em Journal of the ACM}, 49:672--713, 2002.

\bibitem{Basin17eve}
David~A. Basin, Hans Gersbach, Akaki Mamageishvili, Lara Schmid, and Oriol
  Tejada.
\newblock Election security and economics: It's all about {Eve}.
\newblock In {\em Proceedings of {E-Vote-ID}}, pages 1--20, 2017.

\bibitem{Basin16humanerrors}
David~A. Basin, Sasa Radomirovic, and Lara Schmid.
\newblock Modeling human errors in security protocols.
\newblock In {\em Computer Security Foundations Symposium, {CSF}}, pages
  325--340. {IEEE} Computer Society, 2016.

\bibitem{Behrmann04uppaal-tutorial}
G.~Behrmann, A.~David, and K.G. Larsen.
\newblock A tutorial on {\sc uppaal}.
\newblock In {\em Formal Methods for the Design of Real-Time Systems:
  {SFM-RT}}, number 3185 in LNCS, pages 200--236. Springer, 2004.

\bibitem{Bella14concertina}
Giampaolo Bella, Paul Curzon, Rosario Giustolisi, and Gabriele Lenzini.
\newblock A socio-technical methodology for the security and privacy analysis
  of services.
\newblock In {\em {COMPSAC} Workshops}, pages 401--406. {IEEE} Computer
  Society, 2014.

\bibitem{Bella15servicesecurity}
Giampaolo Bella, Paul Curzon, and Gabriele Lenzini.
\newblock Service security and privacy as a socio-technical problem.
\newblock {\em J. Comput. Secur.}, 23(5):563--585, 2015.

\bibitem{Benaloh94receipt}
J.~Benaloh and D.~Tuinstra.
\newblock Receipt-free secret-ballot elections.
\newblock In {\em Proceedings of the twenty-sixth annual ACM symposium on
  Theory of Computing}, pages 544--553. ACM, 1994.

\bibitem{Bourne70concepts}
L.~E. Bourne.
\newblock Knowing and using concepts.
\newblock {\em Psychol. Rev.}, 77:546--556, 1970.

\bibitem{Buldas07evoting}
Ahto Buldas and Triinu M{\"a}gi.
\newblock Practical security analysis of e-voting systems.
\newblock In {\em Proceedings of IWSEC}, volume 4752 of {\em Lecture Notes in
  Computer Science}, pages 320--335. Springer, 2007.

\bibitem{Carlos12ceremonies}
Marcelo~Carlomagno Carlos, Jean~Everson Martina, Geraint Price, and
  Ricardo~Felipe Cust{\'{o}}dio.
\newblock A proposed framework for analysing security ceremonies.
\newblock In {\em {SECRYPT}}, pages 440--445. SciTePress, 2012.

\bibitem{Chatterjee10strategylogic}
K.~Chatterjee, T.A. Henzinger, and N.~Piterman.
\newblock Strategy {L}ogic.
\newblock {\em Information and Computation}, 208(6):677--693, 2010.

\bibitem{Cortier16sok-verifiability}
V.~Cortier, D.~Galindo, R.~K{\"{u}}sters, J.~M{\"{u}}ller, and T.~Truderung.
\newblock {SoK}: Verifiability notions for e-voting protocols.
\newblock In {\em {IEEE} Symposium on Security and Privacy}, pages 779--798,
  2016.

\bibitem{Culnane15vvote}
C.~Culnane, P.Y.A. Ryan, S.A. Schneider, and V.~Teague.
\newblock vvote: {A} verifiable voting system.
\newblock {\em {ACM} Trans. Inf. Syst. Secur.}, 18(1):3:1--3:30, 2015.

\bibitem{Culnane16benalohGT}
Chris Culnane and Vanessa Teague.
\newblock Strategies for voter-initiated election audits.
\newblock In {\em Decision and Game Theory for Security: Proceedings of
  GameSec}, volume 9996 of {\em Lecture Notes in Computer Science}, pages
  235--247. Springer, 2016.

\bibitem{David15sociotechnical-attacks}
Nicolas David, Alexandre David, Ren{\'{e}}~Rydhof Hansen, Kim~Guldstrand
  Larsen, Axel Legay, Mads~Chr. Olesen, and Christian~W. Probst.
\newblock Modelling social-technical attacks with timed automata.
\newblock In {\em Proceedings of International Workshop on Managing Insider
  Security Threats, {MIST}}, pages 21--28. {ACM}, 2015.

\bibitem{Davis15commonsense}
E.~Davis and G.~Marcus.
\newblock Commonsense reasoning.
\newblock {\em Communications of the ACM}, 58(9):92--103, 2015.

\bibitem{Delaune06coercion}
S.~Delaune, S.~Kremer, and M.~Ryan.
\newblock Coercion-resistance and receipt-freeness in electronic voting.
\newblock In {\em Computer Security Foundations Workshop, 2006. 19th IEEE},
  pages 12--pp. IEEE, 2006.

\bibitem{Fagin95knowledge}
R.~Fagin, J.~Y. Halpern, Y.~Moses, and M.~Y. Vardi.
\newblock {\em Reasoning about Knowledge}.
\newblock MIT Press, 1995.

\bibitem{Feldman00conceptlearning}
J.~Feldman.
\newblock Minimization of {Boolean} complexity in human concept learning.
\newblock {\em Nature}, 407:630--3, 11 2000.

\bibitem{Ghallab04planning}
M.~Ghallab, D.~Nau, and P.~Traverso.
\newblock {\em Automated Planning: Theory and Practice}.
\newblock Morgan Kaufmann, 2004.

\bibitem{Hunker11insider-threats}
Jeffrey Hunker and Christian~W. Probst.
\newblock Insiders and insider threats - an overview of definitions and
  mitigation techniques.
\newblock {\em J. Wirel. Mob. Networks Ubiquitous Comput. Dependable Appl.},
  2(1):4--27, 2011.

\bibitem{Jamroga18Selene}
W.~Jamroga, M.~Knapik, and D.~Kurpiewski.
\newblock Model checking the {SELENE} e-voting protocol in multi-agent logics.
\newblock In {\em Proceedings of the 3rd International Joint Conference on
  Electronic Voting (E-VOTE-ID)}, volume 11143 of {\em Lecture Notes in
  Computer Science}, pages 100--116. Springer, 2018.

\bibitem{Jamroga17preventing}
W.~Jamroga and M.~Tabatabaei.
\newblock Preventing coercion in e-voting: Be open and commit.
\newblock In {\em Electronic Voting: Proceedings of E-Vote-ID 2016}, volume
  10141 of {\em Lecture Notes in Computer Science}, pages 1--17. Springer,
  2017.

\bibitem{Jamroga19natstrat-aij}
Wojciech Jamroga, Vadim Malvone, and Aniello Murano.
\newblock Natural strategic ability.
\newblock {\em Artificial Intelligence}, 277, 2019.

\bibitem{Jamroga19natstratii}
Wojciech Jamroga, Vadim Malvone, and Aniello Murano.
\newblock Natural strategic ability under imperfect information.
\newblock In {\em Proceedings of the 18th International Conference on
  Autonomous Agents and Multiagent Systems AAMAS 2019}, pages 962--970.
  IFAAMAS, 2019.

\bibitem{Juels05coercion}
A.~Juels, D.~Catalano, and M.~Jakobsson.
\newblock Coercion-resistant electronic elections.
\newblock In {\em Proceedings of the 2005 ACM workshop on Privacy in the
  electronic society}, pages 61--70. ACM, 2005.

\bibitem{Kusters10game}
R.~K{\"u}sters, T.~Truderung, and A.~Vogt.
\newblock A game-based definition of coercion-resistance and its applications.
\newblock In {\em Proceedings of the 2010 23rd IEEE Computer Security
  Foundations Symposium}, pages 122--136. IEEE Computer Society, 2010.

\bibitem{Martimiano15ceremony}
T.~Martimiano, E.~Dos Santos, M.~Olembo, and J.E. Martina.
\newblock Ceremony analysis meets verifiable voting: Individual verifiability
  in {Helios}.
\newblock In {\em SECURWARE}, 2015.

\bibitem{Martimiano16threatmodelling}
Taciane Martimiano and Jean~Everson Martina.
\newblock Threat modelling service security as a security ceremony.
\newblock In {\em 11th International Conference on Availability, Reliability
  and Security, {ARES}}, pages 195--204. {IEEE} Computer Society, 2016.

\bibitem{Mogavero14behavioral}
F.~Mogavero, A.~Murano, G.~Perelli, and M.Y. Vardi.
\newblock Reasoning about strategies: On the model-checking problem.
\newblock {\em ACM Transactions on Computational Logic}, 15(4):1--42, 2014.

\bibitem{Ryan15verifiability}
Peter Y.~A. Ryan, Steve~A. Schneider, and Vanessa Teague.
\newblock End-to-end verifiability in voting systems, from theory to practice.
\newblock {\em {IEEE} Security {\&} Privacy}, 13(3):59--62, 2015.

\bibitem{Ryan10atemyvote}
P.Y.A. Ryan.
\newblock The computer ate my vote.
\newblock In {\em Formal Methods: State of the Art and New Directions}, pages
  147--184. Springer, 2010.

\bibitem{Santos18phd}
F.P. Santos.
\newblock {\em Dynamics of Reputation and the Self-organization of
  Cooperation}.
\newblock PhD thesis, University of Lisbon, 2018.

\bibitem{Santos18SocialNormComplexity}
F.P. Santos, F.C. Santos, and J.M. Pacheco.
\newblock Social norm complexity and past reputations in the evolution of
  cooperation.
\newblock {\em Nature}, 555:242--245, 2018.

\bibitem{Shoham09MAS}
Y.~Shoham and K.~Leyton{-}Brown.
\newblock {\em Multiagent Systems - Algorithmic, Game-Theoretic, and Logical
  Foundations}.
\newblock Cambridge University Press, 2009.

\bibitem{Tabatabaei16expressing}
M.~Tabatabaei, W.~Jamroga, and Peter Y.~A. Ryan.
\newblock Expressing receipt-freeness and coercion-resistance in logics of
  strategic ability: Preliminary attempt.
\newblock In {\em Proceedings of the 1st International Workshop on {AI} for
  Privacy and Security, PrAISe@ECAI 2016}, pages 1:1--1:8. {ACM}, 2016.

\bibitem{Verifiedvoting19policy}
Verified Voting.
\newblock Policy on direct recording electronic voting machines and ballot
  marking devices.
\newblock 2019.

\end{thebibliography}

%
%
%
%
%
%

\end{document}